# Small Groups of Galaxies as Cosmological Clocks

Paolo Tozzi

*II Università di Roma, v. Ricerca Scientifica 1, 00133 Roma, Italy*

Fabio Governato

*Dept. of Astronomy, University of Washington, Seattle WA 98195, USA*

Alfonso Cavaliere

*II Università di Roma, v. Ricerca Scientifica 1, 00133 Roma, Italy*

**Abstract.**
We study the formation and the evolution of small groups of galaxies using new Monte Carlo simulations. These are directly based on the random walk approach to the statistics of condensations collapsing by gravitational instability, and the results are consistent with the Press & Schechter formula. We stress how observational features of groups, such as the galaxy membership, sensitively depend on the global dynamics of the universe. This is because the amount of aggregations between infalling galaxies is governed by the ratio of the group age to the time scale for dynamical friction; this ratio in turn strongly depends on cosmology. Thus in low–density universes - even if flat - the aggregations would lead often to a strongly dominant merger, contrary to the observations of compact groups.

## 1. Introduction

We submit that the galaxy population of small groups of galaxies ($M \approx 10^{13}\, M_\odot$) constitutes a sensitive probe of the global dynamics of the universe.

To construct a realistic model for the evolution of such a population, we start from the results of Governato, Tozzi & Cavaliere (1996, hereafter GTC). There we showed with N–body simulations that the structure of small groups results from the interplay of three basic processes: collapse of the initial density perturbation, aggregations between the member galaxies, and secondary infall of additional galaxies. Our main result was that in a critical universe the "merging runaway" (see Cavaliere, Colafrancesco & Menci 1991) is over by the time when secondary infall is still on, so that the galaxy population of the groups is effectively rejuvenated. This appears to be a key to understand the very presence today of compact groups (Hickson 1982).

On the other hand, in open universes the suppression of secondary infall leaves the dominant role to aggregations between galaxies, which lead to a single merger remnant (see also Tozzi, Governato & Cavaliere 1996).

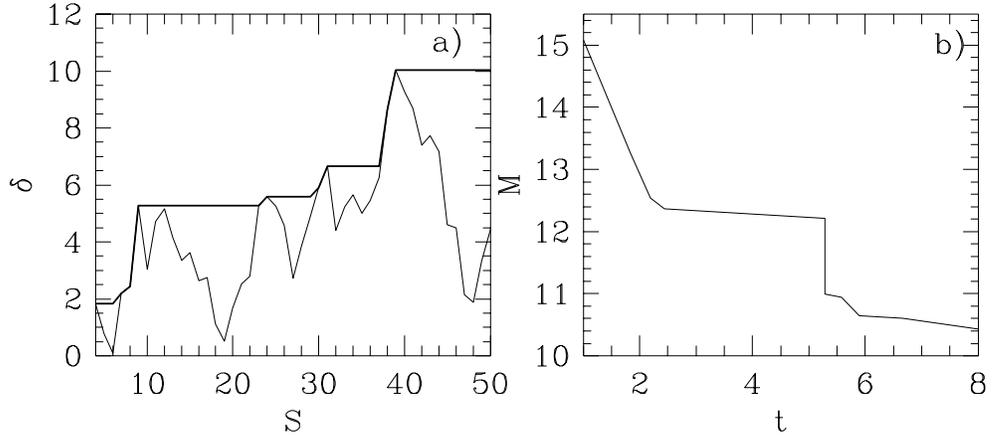

Figure 1. a) Random walk in the $\delta - S$ plane; b) the corresponding mass history in the $M - t$ plane (arbitrary units).

Here we intend to examine further the role of aggregations in small groups of galaxies for a wider range of cosmologies and of power spectra for the initial perturbations. For the sake of clarity we refer by "aggregation" to the coalescence of the baryonic components of the galaxies, as distinct from "merging" of dark halos.

## 2. Mass histories from random walking density perturbations

We use an efficient way to compute a large number of mass histories of groups, constituted by Monte Carlo simulations. There is no unique way to generate such mass histories. Lacey & Cole (1993) proposed a Monte Carlo which "tends to overestimate halo ages" relative to N–body simulations. Kauffman & White (1993) constructed merger trees by extracting *a posteriori* the progenitor mass distribution from the analytical formula (Bond et al. 1991, Bower 1991).

We propose a *new* version *directly* based on the random walk description of the statistics of the perturbation field, as proposed by Bond et al. (1991). We start directly from the random walks (trajectories) in the plane $\delta - S$, $\delta$ being the density contrast at a resolution $S(M)$ given by the variance of the linear field in the form $S(M) = \sigma^2(M)$. The increment in $\delta$ after a step $dS$ is a stochastic variable extracted from a Gaussian with variance $dS$. Once $\delta$ first exceeds a given threshold, an object appears at the "time" $\delta(z)$ at the variance $S(M)$ (see fig 1a,b). For every merging event, i.e. for every discrete variation in the mass of the group, a new branch appears in order to conserve mass, so that a complete merger tree can be constructed. Our merger trees are tested for consistency with the analytic mass distribution of the progenitors (Bond et al. 1991, Bower 1991).

We stress that, at variance with previous simulations, our Monte Carlo is based on a true Gaussian process. Moreover, merging events univocally correspond to a discontinuity in the mass history $M(z)$ (i.e., they do not depend on $d\delta$, see fig 1a). We also stress that this procedure is consistent with the Press &

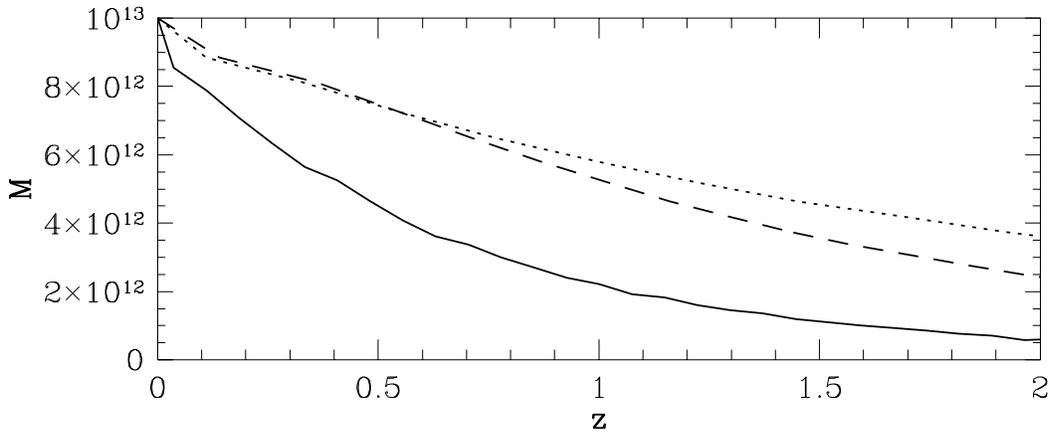

Figure 2. Average mass histories of the main progenitor. Solid line: TCDM; dotted line: OCDM; dashed line: $\Lambda$CDM.

Schechter (1974) formula; in particular, the same dispersion of $M - z$ is obtained with the only requirement of mass conservation in the merger tree.

Next we present an application to the evolution of groups. For simplicity, we follow only the most massive branch of each merger tree, i.e. the "main progenitor".

### 3. Histories of small groups of galaxies

For the power spectrum of the initial perturbations, we select three representative CDM models: the tilted CDM model in the critical universe (TCDM), the open CDM with $\Omega_0 = 0.2$ (OCDM), and finally the flat–low density CDM model ($\Lambda$CDM) with $\Omega_0 = 0.2$ and $\Lambda = 0.8$ (Bardeen et al. 1986, White et al. 1996). The normalization of the power spectra, expressed in terms of $\sigma_8$ (the rms value of the linear fluctuations on scales $8\,h^{-1}$ Mpc) are calibrated using the local observations. Among them most solid are the local abundance and the temperature distribution of clusters (White, Efstathiou & Frenk 1993, Henry & Arnaud 1992). These values are: $\sigma_8 = 0.6$ for TCDM, $\sigma_8 = 1$ for OCDM and $\sigma_8 = 1.4$ for $\Lambda$CDM. These are in agreement with the normalization given by COBE for such spectra, within the statistical uncertainties (Bunn & White 1995). Moreover, we take $H_0 = 50$ km sec$^{-1}$ Mpc$^{-1}$ for TCDM and $H_0 = 65$ km sec$^{-1}$ Mpc$^{-1}$ for OCDM and $\Lambda$CDM.

Following the most massive branch of each tree, we compute the average history of the main progenitor (see fig 2). Next, we compute the distribution of formation epochs $z_f$, defined as the redshift at which the main progenitor reaches one half of the final mass; equivalently, we present the distribution of the ages of the groups counted from $z_f$ (see fig 3).

In the critical cosmology present–day halos were assembled only recently, gaining most of their mass by infall of dark matter halos in the fairly recent past. The formation epochs are peaked around $z_f \simeq 0.5$, and about the 85 %

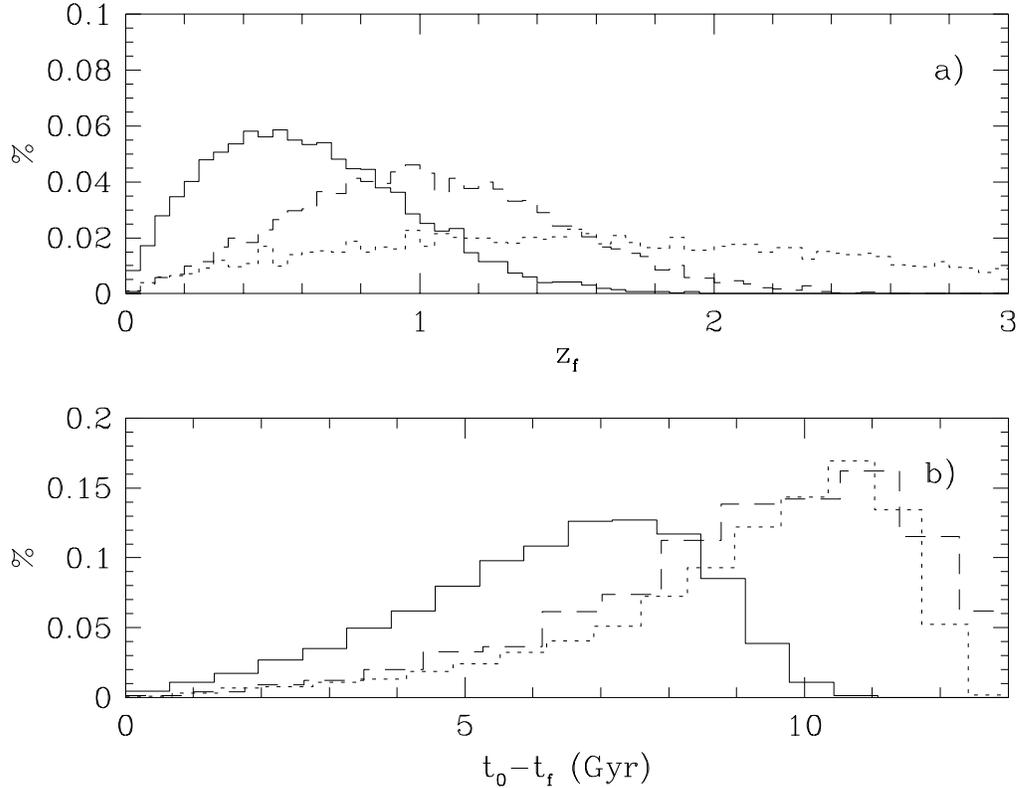

Figure 3. a) Distribution of the formation redshifts $z_f$. b) Ages of the main progenitor in Gyr.

of the present–day population is younger than $z \simeq 1$, i.e., younger than about 8 Gyr.

In OCDM the majority of the merging events, and so most of the mass growth of the halos, occur at high redshifts. In this case the values of $z_f$ are widely distributed around $z_f \simeq 1.5$; more than 80% of the groups are older than 8 Gyr. In $\Lambda$CDM the distribution is intermediate, with average $z_f \simeq 1$.

It is then clear that in the critical case the galaxy population of the groups is *rejuvenated* at recent epochs by the galaxy included in the infalling matter. There is little time for these galaxies to suffer extensive dynamical friction, and so *low* probability to aggregate with the merger remnant.

In OCDM and in $\Lambda$CDM, on the other hand, the galaxy population is formed at earlier epochs. So the majority of the galaxies spend most of their life in the group with enhanced probability to suffer aggregations.

In fact, the time scale which governs the onset of aggregations inside groups is the dynamical friction time scale $T_f$. The classic formula for $T_f$ (see Binney & Tremaine 1987), which is in agreement with the behavior seen in our N–body simulations (GTC), is valid when the infalling galaxies have masses much smaller than the receiving group, a condition satisfied by the vast majority of the infalling halos. If we assume (following Lacey & Cole 1993) that on infalling each halo is stripped of its dark matter component, we obtain a lower bound to the number of accreted baryonic cores, since the infalling halos can host more than one baryonic core (with less dynamical friction).

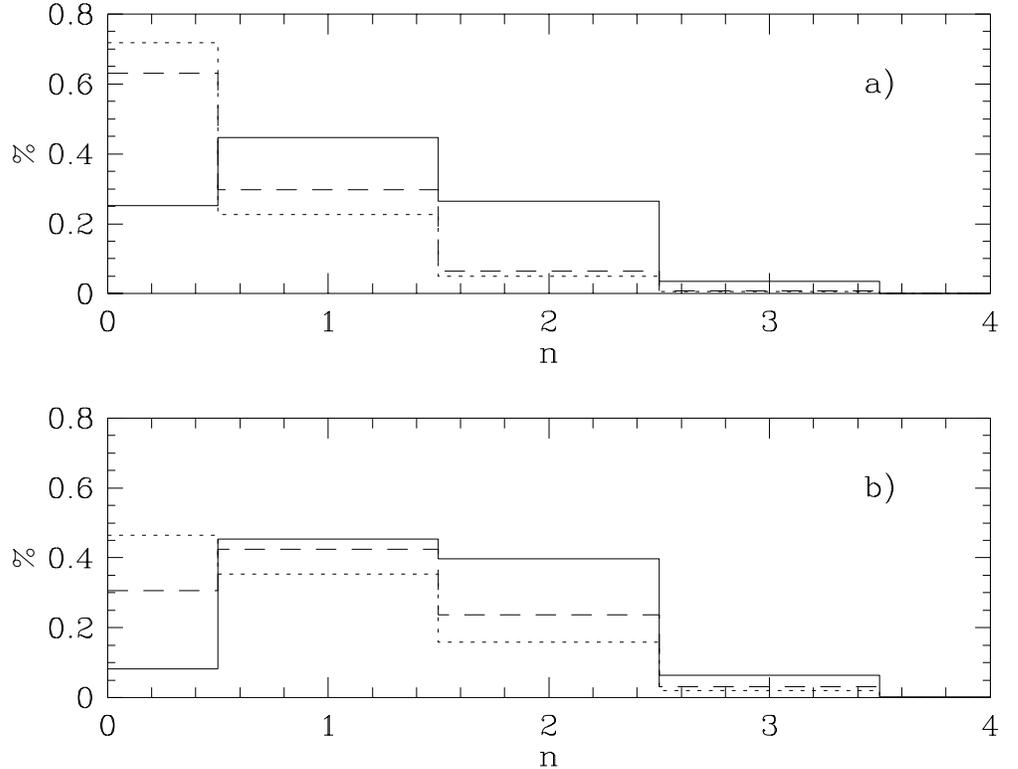

Figure 4. Fraction of present–day groups with $n$ surviving galaxies in different cosmological models (the lines are the same as in fig 2): a) for elongated orbits of the infallen galaxies; b) for circular orbits.

To give an example, we consider halos infalling into the main progenitor for $z < z_f$, and we calculate $T_f$ for halos with $M \geq 10^{12}\ M_\odot$. One such event is interpreted as one bright galaxy infalling into the group, assuming that each halo contains a galaxy with $M_g = f_b M$, where $f_b = \Omega_b/\Omega_0$ is the baryonic fraction (here we assume $\Omega_b = 0.013\ h^{-2}$ consistent with nucleosynthesis, see Walker et al. 1991). The ratio of the group age to $T_f$ sets the amount of aggregations of galaxies with the central merger remnant. In fact, if $T_f$ is smaller than the look-back time to the infall event, the galaxy may be assumed to suffer aggregation with the central merger remnant. In this way we obtain a lower bound to the number of survivors. The result depends on the elongation of the orbits (see GTC). For elongated orbits and TCDM, we see that more than 75% of groups contains at least a bright surviving galaxy, while in OCDM less than 30% of groups contain a bright surviving galaxy. In $\Lambda$CDM the fraction is raised to 37%. For circular orbits, which minimize the dynamical friction, there is still difference between the models (see fig 4). In this sense the groups behaves like cosmological clocks (see fig 5).

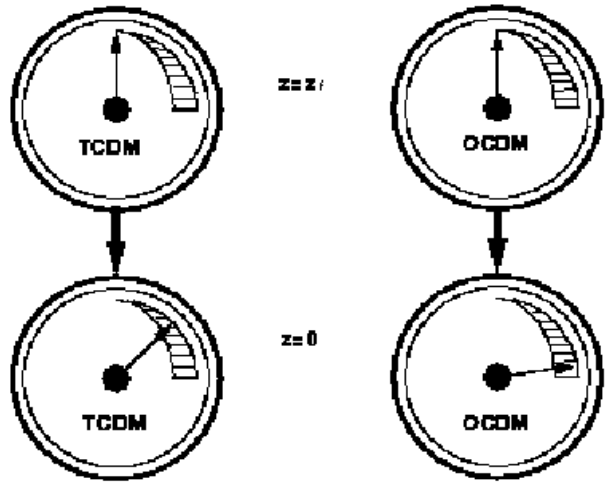

Figure 5. Small groups as $\Omega_0$–clocks. The arrows indicate cosmic time measured in units of $T_f$ (represented with the dashed area), to visualize that with TCDM, only a few units of $T_f$ have elapsed since $z = z_f$; with OCDM instead, the present–day groups are many units of $T_f$ old, and the aggregation process goes to its completion.

## 4. Conclusions

We have studied the link between the structure of small groups of galaxies and cosmology. We stress that such link stems from the interplay of two basic time scales: the age of the small groups (set by the cosmological model) and $T_f$ for the infalling galaxies. These time scales are comparable for small groups of galaxies with $M \sim 10^{13}\ M_\odot$.

The groups constitute sensitive cosmological clocks because their structure is affected *twice* by the cosmological model: as for the distribution of formation epochs, and as for the degree of aggregation between infallen galaxies.

To examine these processes, we have presented a *new*, efficient Monte Carlo scheme to extract mass history of halos directly from the statistics underlying the P&S formula.

Our preliminary results predict that in the critical universe infallen galaxies have a good probability to *escape* aggregations down to the present time, and so constitute true groups of few to several members. In low–density universes, in which the majority of small groups form earlier and afterwards experience little secondary infall, the galaxy population is not rejuvenated; aggregations constitute the important dynamical process which goes to its *completion* and forms a single remnant or a dominant merger surrounded by satellites. Flat low–density universes constitute intermediate cases, but closer to the open version.

We thank N. Menci for many useful discussions, and P. Padovani for his help. We acknowledge research grants from MURST and ASI.